\documentclass[12pt]{iopart}
\usepackage{enumerate}
\usepackage{amssymb}
\usepackage{algorithm}
\usepackage{algorithmic}
\begin{document}

\title[]{Cryptanalysis of the efficient two-party quantum private comparison protocol with decoy photons and two-photon entanglement}
\author{Zhiwei Sun$^1$ and Dongyang Long$^1$}
\address{$^1$School of Information Science and Technology, Sun Yat-sen University, 510006, P.R.China}
\ead{sunzhiwei1986@gmail.com and issldy@mail.sysu.edu.cn}
\begin{abstract}
We analyze the security of the efficient two-party quantum private comparison protocol with decoy photons and two-photon entanglement. It is shown that the compromised third party (TP) can obtain the final comparison result without introducing any detectable errors in the standard semi-honest model. The attack strategy is discussed in detail and an improvement of this protocol is demonstrated. The idea of our attack might be instructive for the cryptanalysis of quantum cryptographic schemes.
\end{abstract}

\maketitle


Protocols for private comparison allow two players holding secret nonnegative integers $x$ and $y$ to compute whether their secret $x$ and $y$ are equal or not without revealing to the other player any additional information about their respective value, which is a variant of the millionaires' problem (also called secure multi-party computation), introduced by Yao \cite{1982Yao,1986Yao}. It has wide applications, such as private bidding and auctions, secret ballot elections, commercial business and data mining. Because of its importance and wide applicability, many protocols for private comparison have been proposed in modern cryptography \cite{1996Fagin,1999Cachin,2001 boudot}. However, in quantum cryptography, Lo \cite{1997Lo} shows that the equality function cannot be securely computed between two players. However, if some additional assumptions are made, the goal can be achieved.

With the help of a third party, Yang et al. proposed the first efficient protocol for quantum private comparison (QPC) based on decoy photons and two-photon entanglement in 2009 \cite{2009Yang}. They proved that their protocol can ensure fairness, efficiency and security in the standard semi-honest (also know as honest-but-curious) model where players are assumed to follow the protocol but may then try to learn additional information from the protocol execution. Here, we study the cryptanalysis of this protocol and we will show that the compromised TP can learn the final comparison result even if he follows the protocol honestly. And a simple solution to this problem is proposed so that it can withstand this kind of attack. For simplicity we will call this protocol YW protocol in the following text.

We first describe YW protocol briefly. Suppose there are a third party TP, and two players Bob and Charlie. The secret inputs of Bob and Charlie are $x$ and $y$, respectively. And the protocol is executed in the standard semi-honest model. The YW protocol consists of the following steps.

(1) The TP, Bob and Charlie agree that the four unitary operations $U_{00} = I$, $U_{01} = \sigma_{z}$, $U_{10} = \sigma_{x}$ and $U_{11} = i\sigma_{y}$ represent two-bit information $00$, $01$,$10$ and $11$, respectively.

(2) The TP randomly prepares a sequence of $n (n > M/2)$ ordered EPR pairs T and the initial states of EPR pairs only known to him. The EPR pairs in the sequence T can be formed two ordered photon sequences $T_{B} = \{t_{B}^{1}, {t_{B}^{2}}, \cdots, {t_{B}^{n}}\}$ and $T_{C} = \{t_{C}^{1}, {t_{C}^{2}}, \cdots, {t_{C}^{n}}\}$. Then the TP adopts the decoy photon technique for preventing the dishonest player from eavesdropping, and sends the sequence $T_{B}$ to Bob and $T_{C}$ to Charlie.

(3) After Bob and Charlie have received all the particles, the TP, Bob and Charlie will check whether the particles are eavesdropped during the transmission by the decoy photon technique. If the error rate is higher than the threshold, then the protocol will be aborted. Otherwise, they proceed to step (4).

(4)Bob and Charlie encode their secrets' hash values $H(X)$ and $H(Y)$ on the photons held by them with the four unitary operations, respectively. Here, $X$ and $Y$ are the binary representations of $x$ and $y$ in $F_{2^{N}}$. To check whether the TP will cheat in the following announcement of his measurement outcomes, Bob and Charlie perform the following two actions \cite{2010Yang}.
\begin{enumerate}
\item They firstly require the TP to publish the initial states of the remaining intact EPR pairs in the original order he initially prepared. The set of these sampling EPR pairs will be used to detect whether the TP is cheating.
\item Then Bob (Charlie) inserts the sampling EPR photons into the encoded photon sequence at the positions determined by the secret random value of $l$ which is generated by using the QKD method and is unknown to the TP. The above operation corresponds to a secret permutation operation applied on the entire photon sequence. We denote the new sequence $T^{'}$.
   \end{enumerate}
For checking eavesdropping in quantum channels, Bob and Charlie send the sequence $T^{'}$ back to the TP by using the decoy photon technique.

(5) After the TP has received all the particles, the TP, Bob and Charlie will first check the eavesdropping during the transmission by the decoy photon technique. If they confirm that there is no eavesdropping, then the TP measures each two correlated photons received form Bob and Charlie with respect to the orthonormal basis $\{|\phi^{\pm}\rangle, |\psi^{\pm}\rangle \}$ and publishes the measurement outcomes. Moreover, he still publishes the initial states of EPR pairs in the original order he initially prepared. Then Bob and Charlie check whether the TP is honest by comparing the measurement outcomes with the TP's beforehand announcements in step (4). If the inconsistency rate between them is higher than the threshold, Bob and Charlie can find the TP is cheating and abort the protocol. Otherwise, they can distill the outcomes of the combination of the unitary operations performed by them from the TP' measurement outcomes and their initial states with error correction and privacy amplification and deduce the comparison result.

The fairness, security and efficiency of the YW protocol have been proven in the original paper. And the fairness means that one player knows the compared result if and only if the other one knows the result; the security of the protocol for the two players means that any information about their secret inputs will not leak including the final comparison result; here, efficiency means the involvement of the TP will improve the efficiency of the protocol compared with the case without the TP.
When we analyze the security of the YW protocol, we will also consider three possible cases as mentioned in Ref. \cite{2009Yang}: (1) the honest TP and a player (Charlie), and a dishonest one Bob. (2) a compromised TP and two honest players Bob and Charlie. (3) the TP colludes with a dishonest player, say Bob. In the standard semi-honest (also known as honest-but-curious) model, parties are assumed to follow the protocol but may then try to learn additional information from the protocol execution. The authors of Ref. \cite{2009Yang,2010Yang} believed that "the TP cannot learn any information about the players' respective private inputs and even about the comparison result in that the TP cannot discriminate between the encoded EPR pairs and the sampling ones because Bob and Charlie insert the sampling ones into the encoded EPR particle sequences at the positions determined by the random number $l$ only known to Bob and Charlie."
However, this conclusion will be broken in the standard semi-honest model. As the compromised TP can learn some additional information from the protocol execution (to be precise, the TP can record the number of each secret EPR pairs), and then follow the protocol honestly; after the end of the protocol, the TP can deduce the comparison result by comparing the number of each of the measurement outcomes (EPR pairs) with the initial recorded number of the EPR pairs. In the following we will describe the TP's eavesdropping in detail.

The TP will record some additional information in step (2) and (5) in YW's protocol \cite{2009Yang}. In step (2), we assume the TP prepares a sequence of $n$ ordered EPR pairs $T$ each randomly in one of the four Bell states only known to him. Additionally, the TP records the number of each Bell states. Here, suppose there are $n_{1}$ $|\phi^{+}\rangle$,  $n_{2}$ $|\phi^{-}\rangle$,  $n_{3}$ $|\psi^{+}\rangle$ and $n_{4}$ $|\psi^{-}\rangle$, where $n_{1}+n_{2}+n_{3}+n_{4}=n$.
Then the TP will honestly execute the protocol with Bob and Charlie.
In step (5), the TP takes a Bell-basis measurement on each two correlated photons received from Bob and Charlie, records these measurement outcomes and publishes his initial states of EPR pairs. Additionally, the TP also records the number of these measurement outcomes. Here, suppose there are $n_{1}^{'}$ $|\phi^{+}\rangle$,  $n_{2}^{'}$ $|\phi^{-}\rangle$,  $n_{3}^{'}$ $|\psi^{+}\rangle$ and $n_{4}^{'}$ $|\psi^{-}\rangle$, where $n_{1}^{'}+n_{2}^{'}+n_{3}^{'}+n_{4}^{'}=n$.
After obtaining the additional information $n_{1}, n_{2}, n_{3}, n_{4}, n_{1}^{'}, n_{2}^{'}, n_{3}^{'}, n_{4}^{'}$, the TP can deduce the comparison result. If there are $n_{i}$ and $n_{i}^{'}$, meet the $n_{i} \neq n_{i}^{'}, i\in \{1, 2, 3, 4\}$, then the secret inputs of Bob and Charlie are not equal. However, if all $n_{i} = n_{i}^{'}, i\in \{1, 2, 3, 4\}$, then the TP will obtain nothing. And the TP will not induce any error in both of the situations.

Now we demonstrate the TP's eavesdropping by an example. Suppose that Bob and Charlie have secret inputs $H(X) = 00 01 10 11$ and $H(Y) = 10 11 00 11$, respectively. The TP prepares a sequence of $8$ ordered EPR pairs $T= \{ |\phi^{+}\rangle, |\phi^{+}\rangle, |\psi^{+}\rangle, |\psi^{-}\rangle, |\phi^{-}\rangle, |\psi^{+}\rangle, |\psi^{+}\rangle, |\psi^{-}\rangle \}$ in step (2). And there are $n_{1}=2$ $|\phi^{+}\rangle$,  $n_{2}=1$ $|\phi^{-}\rangle$,  $n_{3}=3$ $|\psi^{+}\rangle$ and $n_{4}=2$ $|\psi^{-}\rangle$. Here, $n_{1}+n_{2}+n_{3}+n_{4}=8$.
According to the YW's protocol, Bob and Charlie encode their secret inputs $H(X) = 00 01 10 11$ and $H(Y) = 10 11 00 11$ on the EPR pairs with the four unitary operations, respectively.
After the TP taking a Bell-basis measurement on each two correlated photons received form Bob and Charlie in step (5), he records the number of each Bell states, i.e., $n_{1}^{'}=1$ $|\phi^{+}\rangle$,  $n_{2}^{'}=1$ $|\phi^{-}\rangle$,  $n_{3}^{'}=4$ $|\psi^{+}\rangle$ and $n_{4}^{'}=2$ $|\psi^{-}\rangle$. As $n_{1} \neq n_{1}^{'}$ and $n_{3} \neq n_{3}^{'}$, the TP deduces that the secret inputs of Bob and Charlie are unequal.

The reason for our attack is available is that (1) the exact number of each EPR pairs is known to the TP, and the sampling Bell states are not changed during the protocol; (2) if the secret inputs of Bob and Charlie are equal, the encoded Bell states are also unchanged after the unitary operations. So if the secret inputs of Bob and Charlie are equal, all the initial Bell states are unchanged. Obviously, the number of each Bell state has not changed. In other words, if there is a Bell state's number has changed after the measurement, the TP can be convinced that the secret inputs are unequal.

To find a way to improve the YW protocol, it is necessary to revisit the essential reason why the YW protocol is susceptible to our attack. In our opinion, the following fact is the main reason. The sampling EPR pairs for checking the TP's cheating has not changed during the protocol. Though the TP cannot discriminate between the encoded EPR pairs, and those for checking cheating and his cheating attack is invalid in the second security check, he can deduce the comparison result by comparing the number of each Bell states between the initial states and these after the measurement.
Therefore, if Bob and Charlie make unitary operations on the sampling EPR pairs according to some secret value generated by the QKD protocol. Then the checking photons will be changed and the number of each EPR pairs also will be changed. With this modification, YW protocol will be secure against the above attack form the TP. 

In conclusion, we present an sample and effective attack on YW protocol \cite{2009Yang,2010Yang}, by which the compromised third party (TP), learning some additional information from the protocol execution, can obtain the comparison result in the standard semi-honest model, and some discussions on how to improve the protocols are presented. Therefore, more attention must be paid to the additional information which may be obtained by eavesdropper in the design and analysis of such protocols. The less additional information leaked, the more secure our protocol designed.

 \ack \indent This work is in part supported by the Key Project of NSFC-Guangdong
Funds (No.U0935002).

\end{document}